# Evaluating Web Search Engines Results for Personalization and User Tracking

Shamma Rashed*, Tasnim Said*, Amal Abdulrahman*, Arsiema Yohannes*, Monther Aldwairi*
*College of Technological Innovations
Zayed University
Abu Dhabi, UAE
{201602375, 201501004, 201602734, m80008144, monther.aldwairi}@zu.ac.ae

*Abstract*—Being contemporary with this technology-driven era, search engines have appeared on the scene to play an irrefutably profound role in our everyday life. Latterly, light has been shed on the trend of personalization, which comes into play whenever different search results are being tailored for a group of users who have issued the same search query. The unpalatable fact that myriads of search results are being manipulated has perturbed a horde of people. With regards to that, personalization can be instrumental in spurring the Filter Bubble effects, which revolves around the inability of certain users to gain access to the typified contents that are allegedly irrelevant per the search engine's algorithm.

In harmony with that, there is a wealth of research on this area. Each of these has relied on using techniques revolving around creating Google accounts that differ in one feature and issuing identical search queries from each account. The search results are often compared to determine whether those results are going to vary per account. Thereupon, we have conducted six experiments that aim to closely inspect and spot the patterns of personalization in search results. In a like manner, we are going to examine how the search results are going to vary accordingly. In all of the tasks, three different metrics are going to be measured, namely, the number of total hits, the first hit, and the correlation between hits. Those experiments are centered around fulfilling the following tasks. Firstly, setting up four VPNs that are located at different geographic locations and comparing the search results with those obtained in the UAE. Secondly, performing the search while logging in and out of a Google account. Thirdly, searching while connecting to different networks: home, phone, and university networks. Fourthly, using different search engines to issue the search queries. Fifthly, using different web browsers to carry out the search process. Finally, creating and training six Google accounts.

## I. INTRODUCTION AND BACKGROUND

Masses of people use search engines for numerous reasons, including discovering authoritative sources, tuning in with the latest breaking news, and settling on purchasing decisions. The returned search results have an eminently significant implications. To cite an instance, the rank of a certain result can have a drastic impact on business outcomes, political elections, and foreign affairs. More than that, search engines are notorious for personalizing the returned results. Even though personalization can retrieve relevant results, it can also raise some concerns about the Filter Bubble effect. This intellectual isolation is centered around returning the results that the algorithm believes the users are interested in, whereas other relevant results remain hidden [1].

A majority of search engines function by building an index that is based on crawling, which is often seen in Google, Yahoo, and countless other engines for them to look for new pages to index. Under those circumstances, special tools that are generally known as crawlers, spiders or bots continuously navigate the web looking to find new pages. A list of website URLs discovered in previous crawls is going to lay the first visible stone in the foundation of the bots' duties. Those duties are fulfilled after detecting new links and adding them to the database of sites that they want to index. Thereupon, the search engine's algorithm is going to display subset of their indexed pages that match the searched terms and is most likely of interest to you. Despite that, some engines can grant you the ability to change the rank of pages [2].

Google's duty circles around figuring out how to match and display the information that is the most germane to whatever the user types from its database. Thereby, Google is under an obligation to unerringly categorize and rearrange the information stored in the database. Along the same line, the search engine is also expected to display the stored information in less than a second. On the grounds of this, Google has surpassed other search engines in the accuracy of its search results. This conspicuous success was a consequence of laying out the groundwork for the page ranking system. In parallel with this, they decided to use backlinks as a proxy dedicated to votes. A page is considered to be more authoritative on a certain topic when it receives a great number of links. Therefore, links are viewed as votes that heighten the validity of pages [3].

Google developed an algorithm that is known as RankBrain. This technology has played a crucial role in generating the results of a certain query. The key to RankBrain's success revolves around the fact that it uses artificial intelligence to constantly improve its performance. In light of this, the newly processed information or search queries are returned with a higher level of accuracy and precision. RankBrain usually looks for two ranking factors namely links and words. With this in mind, the connections between those factors are analyzed by RankBrain for Google to understand the context behind the users' queries [3].

Online users happened to stumble across some patterns of personalization, which was manifested when a group of users searching for the same term were bewildered to receive different results. For instance, searching for "ice cream" in Dubai and Ajman is going to display different results for ice cream parlors that are relevant to the geographical proximity. Ergo, a multitude of people have raised the alarm about how search engines continue to pull the strings in which they have the upper hand in deciding what type of information can be displayed. That being the case, this research is endeavoring to have a grasp of how search engines are manipulating the results and whether any malign intentions were intended [4].

## II. RELATED WORK

There has been a wealth of experimental and theoretical studies on personalization. Each of these has closely examined and measured its patterns.

Hannak et all. [1] worked on a study that was centered on measuring the personalization of web searches and came up with three experiments as follows. First, create three Google accounts. Second, issue the same search query from each account once a day using PhantomJS. For this experiment, "a list of queries that has breadth and impact" was chosen. Then, they compared the query results to check whether they are going to vary per account. They confirmed personalization, because they found that the AMT results have more probability of varying from the "control result than the control results vary from each other". Surprisingly, queries related to politics and companies are usually the most personalized. This is especially relevant when searching for a name of a certain company. Similarly, they also noticed that personalization can be based on geographic location. On the other hand, it had been noted that the least personalized results were found in queries revolving around factual and health-related terms. In the same way, top ranks are usually less personalized than bottom ranks. However, a further critique of this study can be focused on the simplicity of the experiments and lack of dynamic analysis [5].

Speretta and Gauch [8] collaborated on a study about personalizing searches based on histories in which they decided to conduct five experiments. The first experiment investigated the number of training queries required to create a profile, and the number of the profile's concepts that should be used during the process of calculating the similarity between the profile and the document. The second experiment repeated the first experiment with building profiles using snippets, instead of queries. The third experiment used the best conceptual rank found in the first experiment, a total of thirty returning queries and exactly one concept. Furthermore, the value of α is going to be changed every time in the final rank calculation. Similarly, the fourth experiment combined the search engine's rank with the conceptual rank during the process of calculating the result's final rank. Finally, the fifth experiment was conducted to evaluate two unseen queries. Moreover, the conceptual rank is going to be calculated by using the query-based and snippet-based profiles. After that, the conceptual rank will be compared to the original search engine rank.

The researchers found that using a profile built from thirty queries is going to contribute to heightening the rank of the selected result by 33%. On the other hand, using a profile built from snippets of 30 results showed a slight improvement. In the same manner, the snippet-based profile has helped more queries than it has hurt. The best results were found when the conceptual ranking took one concept from the query-based profile and two from the snippet-based profile. They conducted many experiments, and they also made sure to diversify the measured metrics when carrying out each task to obtain precise results. Conversely, the weakness of this study is that the results presentation lacked well-rounded observations and inferences as well as small number of queries [7].

Fernando, Du, and Ashman [8] first experiment compared search engines such as Google, Yahoo, and Bing to a group of private search engines, which are DuckDuckgo, StartPage, and Ixquick. Afterward, they used "Laptop Price" besides "Music Concerts" as their search terms. They signed in to Gmail and switched off the personalization parameter and location spoofing. Moreover, they switched off the personalization parameter to block a certain page content that they searched on so that cookies cannot store data. Furthermore, the location spoofing is switched off because they want to record the results without being influenced by a certain location. The second experiment was the same as the first except that the personalization parameter was switched on while signing in to the Gmail account. The last experiment searched for the same terms and using the same search engines with the location spoofing and personalization parameters turned on. They used Tor in the location spoofing to allow the users to browse anonymously. Consequently, the observed results in the first two experiments had less variance between them. Likewise, while signing in with an account or without, even when the personalization parameter was switched on, off, or kept by default, the results were not greatly varied. On the other hand, a major difference was noticed in the result of the last experiment. For instance, in the first search issued without Tor, the results focused on Australia's context, but when they repeated the search using Tor, they noticed a significant difference between the results. Eventually, those experiments have many pros. The researchers knew about the huge role that cookies play in storing the user's data and the possibility of returning the same results in different experiments. Hence, they used a method to prevent the preservation of data and disabled the cookies.

Robertson, Lazer, and Wilson [9] researched personalization in web searches associated with political terms. They built a custom Chrome extension. This experiment allowed them to automatically maintain SERPs and retrieve it within the participants' browsers in incognito and standard modes. Moreover, the defined list of search queries consisted of terms related to locations, names of people, and groups linked to Donald Trump's opening. They found a variation between the results that they observed in the experiment. As an example, the URLs, pictures, and Google maps were different in each query and for each participant. As a final point, the advantages of conducting the survey and experiment are concentrated on using the participants' browsers and their logins to collect the search data. On the other hand, their work was not diverse and only focused on one experiment.

Dou, Song, and Wen [10] did a study to see whether personalization is effective for different users and different search queries. The found that the strategy of click-based personalization showed an improvement in the performance of the web search. On the other hand, the disadvantages are associated with the limitation of the number of test queries and not considering n-grams [11]

Daoud, Tamine-Lechani, and Boughanem [12] did a study about the effect of using a "concept-based user context for search personalization". The researchers experimented to prove that re-ranking using the concept-based user context makes the results better and more relevant than the basic search. For the experiment, they used two data sets: TREC collection and the ODP data set (they created). They used the "Mercure" search engine to "index the collection of super documents". The researchers found that performing a personalized search is going to improve the retrieved precision of most of the queries.

## III. METHODOLOGY

Our study focuses on measuring the personalization of web searches in which six experiments are going to be conducted to accomplish our main objectives. Under those circumstances, we came up with five terms that cover a wide range of areas. Therefore, we are going to issue different search queries about terms, including "Donald Trump", "makeup", "weapons", "Felicity Huffman", and "Titanic" in all six experiments. We are seeking to fulfill the following. First, setting up four VPNs that are located at different locations namely the United States of America, Germany, China, and Italy. To achieve this, we decided to download a private mobile browser known as "Aloha" and perform the search process in each of the stated countries to check whether the results are going to vary by country. Then, we will compare the results against those obtained in the UAE. After that, we are going to record the first hit and the correlation between hits.

Second, log in to four Gmail accounts to carry out the searching process of the selected terms and look into the number of total hits, the first hit, and the correlation between hits. Afterward, we are going to repeat the second experiment when we are logging out from those four Gmail accounts and look into the same metrics. Subsequently, we are seeking to compare the returned results during every login and logout to determine whether they are going to vary accordingly. Third, connecting to the home, phone, and Zayed University (ZU) networks to search for the chosen terms and examine the three metrics. Next, we are going to set the results side by side to measure the differences between them. Fourth, use different search engines to issue the search queries and record any similarities and differences along with the observed metrics. Ultimately, the search process will be issued using Google, DuckDuckGo, Yahoo, and Bing.

Fifth, using different web browsers to carry out the search process. More importantly, we will document any differences besides the measured metrics, conforming to those web browsers including Safari, Google Chrome, Firefox, Opera, and Brave. Finally, creating and training six Gmail accounts. As a start, we are going to log in to the first account and search for a certain term. Following that, we will log in to the second account and search for a term that is related to the one that was previously used in the first account. The motivation behind this task is to find out whether the returned results are going to be linked to the first term or not. That being the case, the stated experiment is going to be applied to four more accounts.

## IV. RESULTS

In this section, we are going to present the results of our experiments. Moreover, each experiment will be introduced and supported by the results obtained along with the appropriate analysis.

### A. Experiment 1

The first experiment was centered on choosing five terms and observing the returned search results from four different countries. After collecting the obtained search results, we are going to compare them against the results found in the UAE. Opportunely, one of our teammates was in Italy during the spring break, so we decided to take this chance to measure the results obtained in Italy. The search queries were issued without using a VPN while being in Italy to heighten the level of accuracy. We can find that the results obtained in Italy were different than those in the UAE. For instance, the number of hits was different for all five search terms. As displayed in Table 1 and 2, none of the countries have the same number of results for each term. In the same way, the correlation between the links on the first page is different. There is a group of websites that can be found in almost all countries, such as the multilingual online encyclopedia known as Wikipedia. However, the order in which those links are displayed can differ from one country to another, depending on the people's interest in a certain country. Another difference revolved around the content of the first page. In actuality, the websites that were displayed on the first page are not the same. Some links can be found in both countries, but most of the other links are rather different. Likewise, the top stories were also different, even though the searching process was performed in each country at the same time. Despite all of the differences, some aspects are expected to be the same in each country. Speaking of which, the videos displayed on the first page were the same in both countries.

When it comes to the second country, we used a VPN to carry out the same searching process in China. We were surprised to find how similar the results in China were to those obtained in the UAE. The first page in both countries displayed the same links. By way of contrast, the top stories and correlation between links were different in some searches. However, the major difference was in the results of Google maps.

As for the third country, we used the VPN again to search for the same terms, but from Germany. We noticed that the results were different than those in the UAE. Likewise, the correlation between the results was also different. Be that as it may, the same links can still be found in the UAE as Table 1 demonstrates. Furthermore, the biggest difference was in the number of total hits and top stories, especially when each country tends to display its results based on its people's interests.

In the final country, we issued the search queries using a VPN from the USA. Although there are some notable differences between the results returned from both countries, we can still find the same links, but in a different order in the first pages of the two countries as shown in Tables 1 and 5. On the other hand, the major difference lies in the number of total hits and top stories.

TABLE 1. RESULT OF UAE

| UAE | | | | | |
|---|---|---|---|---|---|
| Search term | Donald Trump | Felicity Huffman | Makeup | Titanic | Weapons |
| First hit | Wikipedia | Wikipedia | Makeup.com | RMS Titanic-Wikipedia | List of weapons – Wikipedia |
| Different order? | ✓ | ✓ | ✓ | ✓ | ✓ |
| # of hits in the first page | 12 | 12 | 9 | 10 | 10 |
| # of hits | 822,000,000 | 48,300,000 | 1,490,000,000 | 118,000,000 | 1,680,000,000 |

TABLE 2. RESULT OF ITALY

| Italy | | | | |
|---|---|---|---|---|
| Search term | Donald Trump | Felicity Huffman | Makeup | Titanic |
| First hit | Home \| Donald J. Trump for President | Wikipedia | Makeup.com | RMS Titanic-Wikipedia |
| Different order? | ✓ | ✓ | ✓ | ✓ |
| # of hits in the first page | 14 | 14 | 15 | 14 |
| # of hits | 893,000,000 | 35,300,000 | 14,140,000,000 | 112,000,000 |

TABLE 3. RESULT OF CHINA

| China | | | | | |
|---|---|---|---|---|---|
| Search term | Donald Trump | Felicity Huffman | Makeup | Titanic | Weapons |
| First hit | Twitter | Wikipedia | Makeup.com | RMS Titanic-Wikipedia | List of weapons - Wikipedia |
| Different order? | ✓ | ✓ | ✓ | ✓ | ✓ |
| # of hits in the first page | 14 | 13 | 13 | 10 | 12 |

TABLE 4. RESULT OF GERMANY

| Germany | | | | | |
|---|---|---|---|---|---|
| Search term | Donald Trump | Felicity Huffman | Makeup | Titanic | Weapons |
| First hit | Home \| Donald J. Trump for President | Wikipedia | Makeup \| Sephora | Titanic – Home \| Facebook | Weapons – Wikipedia |
| Different order? | ✓ | ✓ | ✓ | ✓ | ✓ |
| # of hits in the first page | 13 | 13 | 15 | 11 | 12 |

TABLE 5. RESULT OF USA

| USA | | | | | |
|---|---|---|---|---|---|
| Search term | Donald Trump | Felicity Huffman | Makeup | Titanic | Weapons |
| First hit | donaldjtrump.com | Wikipedia | Make-up & cosmetic online kopen bij Douglas | Titanic (1997) - IMDb | List of weapons - Wikipedia |
| Different order? | ✓ | ✓ | ✓ | ✓ | ✓ |
| # of hits in the first page | 15 | 14 | 13 | 12 | 10 |

*B. Experiment 2*

In the second experiment, we compared between the returned search results when we are logging in and out of Gmail. Furthermore, we found the same results in both cases. However, the number of total hits and the correlation between hits was different. Figure 1 demonstrates the number of hits displayed on the first page for each term when we were logging in and out. As the figure suggests, a clear difference can be easily noticed. By way of example, the number of hits that can be found in the first page when we issued the search term "Felicity Huffman" during the log in process was 11 but decreased to 10 when we logged out.

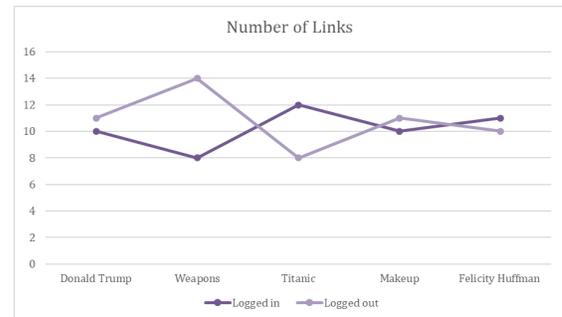

*Fig 1. Number of links displayed on the first page for each search term while logging in and out of Gmail.*

*C. Experiment 3*

The third experiment revolved around issuing the same search terms while being connected to the home, phone, and ZU networks. The number of total hits, correlation between hits and the returned search results were different. There was a diversity in the results returned in each network. Over and above that, the first hit in each network was also different.

*D. Experiment 4*

The fourth experiment relied on performing the searching process using different search engines. Therefore, we issued the search terms on Bing, DuckDuckGo, Yahoo and Google. Furthermore, we noticed many differences. For instance, the correlation between hits was different. Similarly, some of the returned results were different as well. However, our findings regarding the number of total hits can be only looked at in terms of two search engines. The number of total hits was not displayed in neither DuckDuckGo nor Yahoo. Hence, we decided to present our findings on the number of total hits in

Google and Bing on which the number of hits widely differed. Moreover, although there was a diversity in the returned results, some of the results were repeated in other search engines. Table 6 displays the first hit in each search engine. Noticeably, the most common hit was Wikipedia. On top of that, we can easily see that the first hits in Bing are also the first hits on Google and Yahoo. On the other hand, the top results of the term "Titanic" in almost all search engines were about the real-life tragedy of Titanic. Whereas, in Yahoo, the top results were about the film adaptation of Titanic.

*TABLE 6. FIRST HIT FOR EACH SEARCH TERM IN ALL SEARCH ENGINES*

| First Hits | | | | |
|---|---|---|---|---|
| **Search Term** | **DuckDuckGo** | **Google** | **Yahoo!** | **Bing** |
| **Donald Trump** | Donald J. Trump – Official Site | Donald Trump – Wikipedia | Former White House Official – David Stockman – Free Boo | Trump Donald – Donald J Trump |
| **Titanic** | Titanic – History in an Hour Audiobook by Sinead Fitzgibbon | RMS Titanic – Wikipedia | Titanic Comfort Mitte, Berlin – Best Price Guarantee. | RMS Titanic - Wikipedia |
| **Makeup** | Top 5 Eyelash Growth Serums – 2018's The Best Lash | Makeup Products, Tips, Trends & Tutorials \| Makeup.com | Makeup - Almay@ Official Site \| almay.com | Makeup Products, Tips, Trends & Tutorials \| Makeup.com. |
| **Weapons** | Taser Self – Defense Tools – Now Legal in Massachusetts. | List of weapons – Wikipedia | Weapons – Wikipedia | Weapons – Wikipedia. |
| **Felicity Huffman** | Felicity Huffman – IMDb | Felicity Huffman – Wikipedia | Felicity Huffman – IMDb | Felicity Huffman – Wikipedia. |

*E. Experiment 5*

In the fifth experiment, we used different web browsers to carry out the searching process and compare whether the results are going to vary in each browser. We issued the search terms on five browsers, namely, Safari, Google Chrome, Firefox, Opera and Brave. That being the case, we did notice some clear differences between the results found on each browser. Moreover, the first difference was centered on the number of hits. When we were searching for the same terms in different browsers, we can get different number of results. As shown in Figure 2, five search terms were issued, and the number of results was different in most of the browsers for all five terms. Similarly, the correlation between hits was also different. In actuality, the way that the links and sections are displayed on the first page were different in most browsers. Furthermore, the order of the links and the content that can be viewed in the different sections, such as the google map and top stories sections was different as well. As an example, when we were searching for makeup on all five browsers, the order of the news displayed on the top stories section was different. However, we also noticed that some of the links that were returned in certain browsers were nowhere to be seen in the other browsers.

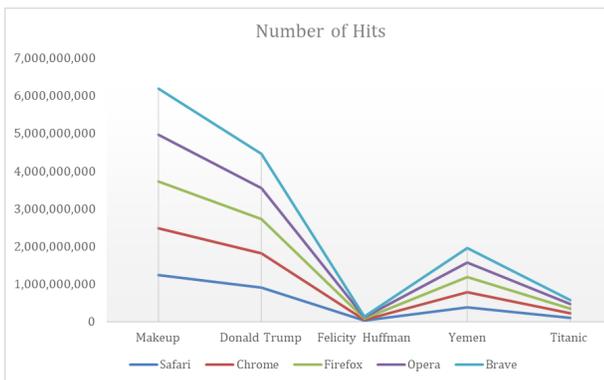

*Fig 2. Returned results in different browsers*

*F. Experiment 6*

As for the final experiment, we decided to create and train six Gmail accounts. To start, we issued a certain search query in the first account and logged out. After that, we logged in to the second account and issued a search term that is related to the previously used term. Under those circumstances, the stated experiment was applied to four more Gmail accounts to find out whether the results of the second account are going to be linked to the term used in the previous search.

In the first account, we issued a search query about the term Donald Trump in which we noticed that most of the returned results were about his scandals. Following that, we logged in to the second account and searched for America. That being the case, we noticed that a lot of results were also about scandals. Despite that, the first hit, number of total hits and correlation between hits were different.

When it comes to the third account, we issued a search query about New Zealand Mosque attack. The returned results related to the devastating attack were displayed. Afterward, we logged in to the fourth Gmail account to search for New Zealand, to our surprise, we found some results about the mosque attack. In fact, some of the links were exactly the same. By way of contrast, the first hit, number of total hits and correlation between hits were rather different.

In the fifth account, we searched for the term Sephora. The displayed results revolved around the official site of Sephora and the location of its stores. Finally, we decided to search for Makeup in which most of the results were about makeup products that are sold on Sephora's website. Having said that, the first hit, number of total hits and correlation between hits were different.

## V. CONCLUDING DISCUSSION

Over the recent years, patterns of personalization were spotted in various internet-based services, especially in web search. Although personalization can be a boon to many users, it can also increase the likelihood of the Filter Bubble effects. On that account, it may prevent many users from accessing certain content.

In this paper, we grappled with this situation by developing a methodology that aimed to measure personalization of web search. Our methodology was centered on conducting six experiments in which we were able to come up with those results. Firstly, the search results, first hit and correlation between hits are going to vary by geographic location. Undoubtedly, search engines tailor their results based on many factors, including search terms that you usually search for, their rank or reputation of those terms, search histories, and the top viral topics in your area [2]. Secondly, the number of total hits and correlation between hits are going to change when we log in and out of a Gmail account. Thirdly, the returned results, first hit, number of total hits and the order of hits are going to differ, depending on the network that we are connected to. Fourthly, different search engines return different results, first hit, number of total hits and order. However, in some cases, a first hit in a certain search engine can be a first hit in another search engine. Fifthly, different web browsers display different results, first hit, number of total hits and order. Finally, logging out of a certain Gmail account after issuing a search query is going to manipulate the results returned in a different Gmail account. As a consequence, Google links the search results returned on the first account to the search results returned on the second account. Be that as it may the first hit, number of total hits, and the correlation between hits are going to be different. We therefore conclude that personalization is caused by issuing search queries from different geographic locations, logging into a Gmail account, logging out from a Gmail account, connecting to different networks, searching on different search engines, and using different web browsers to carry out the searching process.

The next paragraphs are going to address some of the issues that were caused by our work and the directions of our future research.

**Scope**. Our study was on queries that were written in English. Except for the USA, those queries were tested in non-English speaking countries. We leave the inspection of queries in other languages to future work.

**Time**. We measured personalization of web search during a short period. Our future study essays to expand upon our data for a long time.

**Metrics**. In this study, three metrics were measured, namely, number of total hits, first hit, and correlation between hits. We plan on observing more metrics in the future.

**Incompleteness**. Our study only concentrated on some dimensions of personalization in which its absence cannot be claimed. We aim to continue this work by examining the impact of mechanisms that can disable personalization.

**Narrowness**. The methodology that we have developed has studied a narrow aspect of personalization in web search engines. We seek to broaden our research horizons by evaluating the multiple aspects of personalization. (e.g., enabling and disabling cookies, turning search history on and off).